\documentclass[conference]{IEEEtran}
\IEEEoverridecommandlockouts
\usepackage{cite}
\usepackage{amsmath,amssymb,amsfonts}
\usepackage{lipsum,adjustbox}
\usepackage{tikz}
\usepackage{booktabs}
\usepackage{balance}
\usepackage{algorithm,algpseudocode}
\usepackage{graphicx}
\usepackage{textcomp}
\usepackage{xcolor}
\usepackage{url}
\usepackage{hyperref}
\usepackage{breqn}
\usepackage{listings}
\usepackage{multirow}
\usepackage{subcaption}
\newcommand{\subparagraph}{}
\def\BibTeX{{\rm B\kern-.05em{\sc i\kern-.025em b}\kern-.08em
    T\kern-.1667em\lower.7ex\hbox{E}\kern-.125emX}}
\begin{document}

\title{Trust and Believe -- Should We? \\Evaluating the Trustworthiness of Twitter Users
{\footnotesize \textsuperscript	\hfill\href{https://zenodo.org/record/7014109}{\includegraphics[width=2.5\baselineskip,height=3\baselineskip]
		{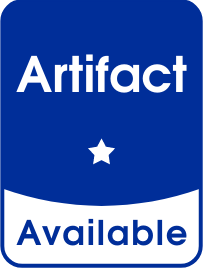}}}
\thanks{This research has received funding from the EU research projects ASCLEPIOS (No. 826093) and CYBELE (No 825355).}
}


\author{\IEEEauthorblockN{1\textsuperscript{st} Tanveer Khan}
\IEEEauthorblockA{\textit{Network and Information Security Group} \\
\textit{Tampere University}\\
Tampere, Finland \\
tanveer.khan@tuni.fi}
\and
\IEEEauthorblockN{2\textsuperscript{nd} Antonis Michalas}
\IEEEauthorblockA{\textit{Network and Information Security Group} \\
\textit{Tampere University}\\
Tampere, Finland \\
antonios.michalas@tuni.fi}}

\maketitle

\begin{abstract}
Social networking and micro-blogging services, such as Twitter, play an important role in sharing digital information. Despite the popularity and usefulness of social media, they are regularly abused by corrupt users. One of these nefarious activities is so-called fake news -- a “virus” that has been spreading rapidly thanks to the hospitable environment provided by social media platforms. The extensive spread of fake news is now becoming a major problem with far-reaching negative repercussions on both individuals and society. Hence, the identification of fake news on social media is a problem of utmost importance that has attracted the interest not only of the research community but most of the big players on both sides - such as Facebook, on the industry side, and political parties
on the societal one. In this work, we create a model through which we hope to be able to offer a solution that will instill trust in social network communities. Our model analyses the behaviour of 50,000 politicians on Twitter and assigns an influence score for each evaluated user based on several collected and analysed features and attributes. Next, we classify political Twitter users as either trustworthy or untrustworthy using random forest and support vector machine classifiers. An active learning model has been used to classify any unlabeled ambiguous records from our dataset. Finally, to measure the performance of the proposed model, we used accuracy as the main evaluation metric.

\end{abstract}


\section{\uppercase{Introduction}}
\label{Introduction}
With one-third of the world's population using some form of social media~\cite{SocialMediaPopulation}, it is evident that the popularity of social networking sites has rapidly increased in recent years. 
This has significantly changed the dynamics of communication across all age groups; the way we work, the way we live, the way we interact with other people and the way we share information have already changed drastically.
Furthermore, social media enables sharing of important information with many people simultaneously, allowing users to reach a bigger audience. 

While social media has its positive sides, it is also important to consider the flip side and properly evaluate its negative impacts. One of the latest negative effects of social media is the so-called fake news phenomenon. It has been proven that the massive distribution of fake news plays an important role in the success or failure of important events and causes~\cite{allcott2017social,metzgar2009social}. Apart from the dissemination and circulation of false information, social networks provide the ideal toolkit for corrupt users to perform a wide range of illegitimate actions such as spamming and political Astroturfing~\cite{wang2010don,grier2010spam}. 

Twitter, with around half a billion users, is one of the three most popular social media platforms. It generates on average~10,000 tweets per second (approximately~500 million tweets per day\footnote{\url{https://www.omnicoreagency.com/twitter-statistics/}})~\cite{alrubaian2017reputation}. It is considered a valuable resource for government agencies, businesses, political parties, financial institutions, fundraising, and many other actors as it enables uncomplicated extraction and dissemination of important information.

A recent study~\cite{hindman2018disinformation} examined~10 million tweets generated by 700,000 different Twitter accounts and linked to~600 fake and conspiracy news sites. It identified clusters of Twitter accounts that linked back to these sites repeatedly, often in ways that seemed coordinated or even automated. In another study, it was found that~6.6 million tweets with fake news were distributed before the 2016 US elections. Different social and political events such as the 2016 US presidential election~\cite{bovet2019influence}  were tainted by a growing number of fake news.

Global concern about the impact of fake news on our societies is on the rise. Hence, there is an immediate need for the design, implementation, and adoption of new systems and algorithms that are able to \textit{identify} and \textit{differentiate} between fake and real news. However, with the increase in the number of social media users\footnote{In 2018, an estimated 2.65 billion people were using social media worldwide, a number projected to increase to almost 3.1 billion in 2021~\cite{SocialMediaPopulation}.}, the quantity of generated content is increasing rapidly, which hinders the identification of fabricated stories~\cite{al2015new} and prevents the identification of a significant amount of information that can potentially give rise to false rumours. Therefore, verifying the credibility of a tweet or assigning a score to users based on the information they have been sharing is a problem that has caught the interest of many academic and industrial researchers~\cite{liu2014tweets,canini2011finding,tinati2012identifying,gupta2012evaluating,moens2014mining,rao2010classifying,al2011experimental,uddin2014understanding}.


\subsection{Our Contribution}
\label{SS:OC}
In this work, we present a model for analysing Twitter users that assigns a score calculated based on their social profiles, tweet credibility and h-index score (i.e. retweets and likes). Users with a higher score are not only considered to be more influential but their tweets are also given greater credibility. Our main contribution can be summarised as follows:

\begin{itemize}

\item First, we generated a dataset of~50,000 Twitter users. For each user, we created a unique profile containing~19 features (discussed in Section~\ref{sec:Methodology}). Our dataset contained only users whose tweets are public and who have friends and followers.

\item For each of the analysed users, we calculated their Social Reputation score (Section~\ref{Methodology:SR}), an h-Index Score (Section~\ref{Methodology:RI}), a Sentiment Score (Section~\ref{Methodology:CNS}), Tweet Credibility (Section~\ref{Methodology:CTS}) and an Influence Score~\ref{Methodology:RS}.

\item Furthermore, we classified each Twitter user account as either trustworthy or untrustworthy. A trustworthy or untrustworthy flag was assigned to each user based on their social reputation, tweet credibility, the sentiment score of a tweet and H-index score of re-tweets and likes, as well as an influence score. 

\item To classify a large pool of unlabeled data, we used an active learning model (a semi-supervised learning algorithm) -- a technique ideal for a situation in which unlabeled data is abundant but manual labeling is expensive~\cite{Settles, Tong}.

\item We measured the performance of our model by using the accuracy metric. This metric measures the percentage of correctly predicted Twitter users~(trustworthy and untrustworthy). 
\end{itemize}

\medskip

We hope that this work will inspire others to perform further research on this emerging problem while at the same time kick-starting a period of greater trust on social media through sustained collaboration between humans and machines.

\subsection{Organisation}
\label{SS:PO}
The rest of this paper is organised as follows: In Section~\ref{sec:Related Work} related work is discussed followed by Section~\ref{sec:Methodology} in which we discuss in detail our proposed approach. The active learning approach and types of classifiers used are discussed in Section~\ref{sec:ActiveLearning}. Section~\ref{sec:Evaluation} features the experimental results and model evaluation and presents the data collection and experimental results of our model. Finally, in Section~\ref{sec:Conclusion}, we conclude the paper.

\section{\uppercase{Related Work}}
\label{sec:Related Work}
Twitter is considered one of the top Online Social Networks (OSNs) that provide a fertile environment for a variety of research purposes. Compared to other popular OSNs, Twitter gains significantly more attention in the research community due to its open policy on data sharing and distinctive features~\cite{ratkiewicz2011detecting}. In 2011, the network had about~175 million unique accounts~\cite{anger2011measuring}, a figure that has grown to an estimated~1.3 billion\footnote{\url{https://www.brandwatch.com/blog/twitter-stats-and-statistics/}}, making it one of the most popular social media platforms. 

Even though openness and vulnerability are two separate issues, there have been many cases where malicious users have taken advantage of Twitter's openness and managed to exploit the service in several ways (e.g. political Astroturfing, spammers sending unsolicited messages, posting malicious links, etc.). 

Despite the important negative impact that the distribution of fake news has on our society, only a handful of techniques for identifying fake news on social media have been proposed~\cite{ratkiewicz2011detecting, shu2017fake, gupta2014tweetcred, wang2010don, grier2010spam}. One of the most popular and promising ideas is to evaluate Twitter users and assign them a credit/reputation score. 

Authors in~\cite{wang2010don} elaborated on the idea that posting duplicate tweets should affect the reputation score of a user since this is a behaviour that legitimate users typically do not engage in. Therefore, posting the same tweet several times would have a negative effect on the user's overall credit score. The authors calculated the edit distance to detect duplication between two tweets posted from the same account. 
Furthermore, the staggering quantities of exchanged messages and information on Twitter have been exploited by users to hijack trending topics~\cite{jain2015hashjacker}. This is a technique used to send unsolicited messages to legitimate users. Additionally, there are Twitter accounts whose only purpose is to artificially boost the popularity of a hashtag with the main aim of increasing its popularity and ultimately making the underlying topic a trend. One BBC report mentioned that \pounds150 was paid on Twitter users to increase the popularity of a hashtag and make it a trend\footnote{\url{https://www.bbc.com/news/blogs-trending-43218939}}.  

To tackle these problems, researchers have used different ways to assess the trustworthiness of tweets and assign an overall rank to users~\cite{gupta2014tweetcred}. Castillo \textit{et al.}~\cite{castillo2011information} measured the credibility of tweets (news topics) based on Twitter features. More precisely, an automated classification technique to detect news from conversational topics was used. Alex Hai Wang~\cite{wang2010don} used followers and friends parameters to calculate the reputation score, which further aided user classification (i.e. to detect spammers). 
%
%
%
Additionally, Saito and Masuda~\cite{kerres2010managing} considered these metrics while assigning a rank to Twitter users. 
In~\cite{gupta2012twitter}, the authors analysed tweets relevant to the Mumbai attacks\footnote{\url{https://www.theguardian.com/world/blog/2011/jul/13/mumbai-blasts}}. 
Their analysis showed that most information providers were unknown while the reputation of the others (based on number of followers) was very low. 
In another study~\cite{gupta2012credibility} that looked at the same event, an information retrieval technique and machine learning algorithm found that only 17\% of the tweets related to the underlying attacks were credible. 

Gilani \textit{et al.}~\cite{gilani2017bots} found that compared to normal users, bots and fake accounts use a large number of external links in their tweets. Hence, analysing other Twitter features such as URLs is of paramount importance for correctly evaluating the overall credibility of a user.
While Twitter has built tools to filter out such URLs, there are several masking techniques that can effectively bypass Twitter's safeguards. 

In this work, we evaluate the trustworthiness and credibility of users~\cite{Michalas:14:StR, Michalas:14:StRM} by analysing a wide range of features (see Table~\ref{tab:FeaturesNotation}). Compared to other works in the area, our model sifts through a plethora of factors that represent signs of possible malicious behaviours and makes honest, fair and precise judgments about the credibility of users with the main aim of engendering community trust.

\section{Methodology}
\label{sec:Methodology}
\begin {figure*}
\begin{adjustbox}{width=\textwidth}
	\resizebox{20cm}{0.6cm}{%
		\begin{tikzpicture}[baseline,thick,scale=1,
		level distance=25mm,
		text depth=.8em,
		text height=2em,
		level 1/.style={sibling distance=4.3cm},
		level 2/.style={sibling distance=1.2cm},
		every node/.style = {transform shape, align=center}]]
		\node {Twitter User \\ Influential Score}
		child { node {Tweet \\ Credibility}
			child { node {\\Original \\ Content \\ Ratio} }
			child { node {URL \\ Ratio} } 
			child { node {Liked \\ Ratio} } 
			child { node {Retweet \\ Ratio} }
			child { node {Hashtag \\ Ratio} } } 	
		child {node {Index \\ Score}
			child { node {Retweet \\ Score} }
			child { node {Like \\ score} }
		}
		child {node {Sentiment \\ Score}
			child { node {Positive \\ Tweets} }
			child { node {Negative \\ Tweets} } 
			child { node {Neutral \\ tweets} } 
		}	
		child { node {Social \\ Reputation} 
			child { node {Follower \\ Count} }
			child { node {Friend \\ Count} }
			child { node {Status \\ Count} }};
		\end{tikzpicture}
	}
\end{adjustbox}
\caption{Twitter User Influence Score Calculation}\label{fig:ReputationDiagram}
\end{figure*}
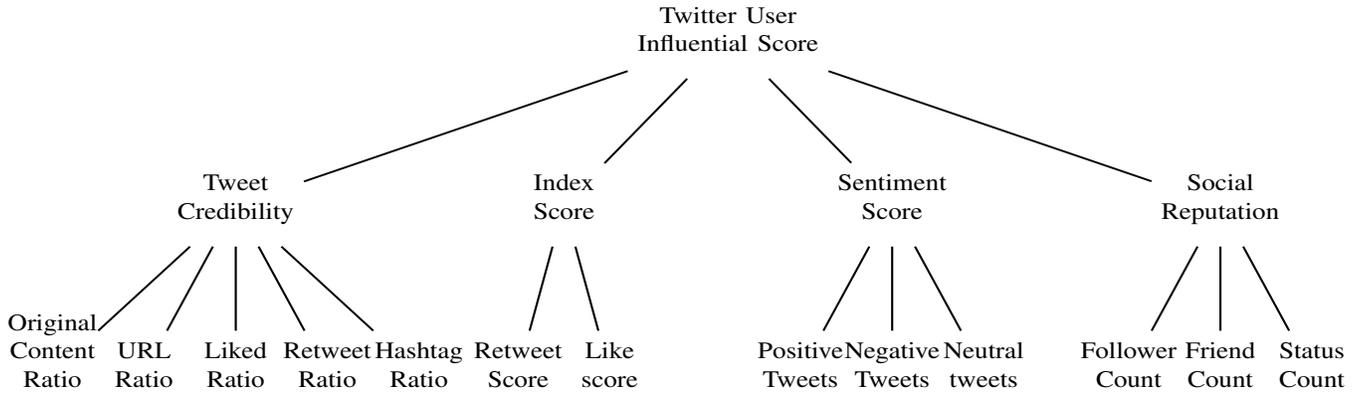

In this section, we discuss our models and main algorithms for calculating the influence score of a user. Our first goal is to help users identify possible information about a political Twitter user by taking into consideration the influence score that results from a proper run of our algorithms. Secondly, political Twitter users are classified either as trustworthy or untrustworthy users based on social reputation, tweet credibility, sentiment score, h-index score and influence score. All those accounts with abusive and harassing tweets, a low social reputation, h-index and influence score are grouped into untrustworthy users while those who are more reputable among users with a high h-index score and more credible tweets as well as high influence score are grouped into trustworthy users.
In the rest of section, we will talk about how to calculate the influence score of a Twitter user. The influence score is calculated by considering both the context and content (tweets) of Twitter accounts. 
In evaluating a user we take into consideration only the Twitter features that can be extracted using \href{https://pypi.org/project/tweepy/}{Twitter API}. Then, we use the outcome of that evaluation and derive more features that help us to provide a more well-rounded and fair evaluation (Section~\ref{SS:DFTU}). 
Figure~\ref{fig:ReputationDiagram} illustrates the main factors we used to calculate user influence scores.

\subsection{Twitter Feature Extraction}
\label{Methodology:TFE}
We now describe in detail all the basic features extracted from Twitter and their importance in the process of assigning a score to each user. 

The key step in assigning a score to Twitter users is to extract the features linked to their accounts. The features can be specific to user accounts such as the number of followers and friends or it can be specific to a user's tweet such as the number of likes, retweets, URLs, etc. In our model, we used these as well as additional features on both a user and content level.

The features used to assign an influence score as well as the relevant notation used throughout the paper is given in table~\ref{tab:FeaturesNotation}.

\begin{table}[!ht]
	\caption{{Selected Attributes for Calculating the Influence Score}}
	\centering
	\begin{tabular}{cccc}
		\toprule
		\hline
		\textbf{Notation} &\multicolumn{1}{c} {\textbf{Description}}\\
		\hline
		\midrule
		\textbf{$SN(u_i)$}: & User screen name\\
		\midrule
		\textbf{$Id(u_i)$}: & User's unique ID\\
		\midrule
		\textbf{$R(u_i)$}: & User influence score\\
		\midrule
		\textbf{$N_{T}(u_i)$}: & Total number of tweets\\
		\midrule
		\textbf{$N_{+ve}(u_i)$}: & Number of neutral tweets\\
		\midrule
		\textbf{$P_{+ve}(u_i)$}: & Number of positive tweets\\
		\midrule
		\textbf{$N_{-ve}(u_i)$}: & Number of negative tweets\\
		\midrule
		\textbf{$S(u_i)$}: & User status\\
		\midrule
		\textbf{$L(u_i)$}: & User list count \\
		\midrule
		\textbf{$R_{hindex}(u_i)$}: & User retweet h-index \\
		\midrule
		\textbf{$L_{hindex}(u_i)$}: & User like h-index \\
		\midrule
		\textbf{$C_{s}(u_i)$}: & User sentiment score \\
		\midrule
		\textbf{$Twt_{cr}(u_i)$}: & User tweet credibility \\
		\midrule
		\textbf{$R_{s}(u_i)$}: & Social reputation score of the user \\
		\midrule
		\textbf{$F_{fol}(u_i)$}: & Number of user followers \\
		\midrule
		\textbf{$F(u_i)$}: & Number of user friends \\
		\midrule
		\textbf{$M(u_i)$}: & Number of user mentions\\
		\midrule
		\textbf{$U_{R}(u_i)$}: & User tweets containing $URLs$\\
		\midrule
		\textbf{$R_{R}(u_i)$}: & User retweet ratio $i$ \\
		\midrule
		\textbf{$L_{R}(u_i)$}: & User liked ratio\\
		\midrule
		\textbf{$O_{R}(u_i)$}: & Original content ratio of the user\\
		\midrule
		\textbf{$H(u_i)$}: & User hashtag ratio\\
		\midrule
		\textbf{$U(u_i)$}: & User URL ratio\\
		\midrule
		\textbf{$M_{R}(u_i)$}: & User mention ratio\\
		\midrule
		\textbf{$R_{n}$}: & Number of retweets\\
		\midrule
		\textbf{$L_{n}$}: & Number of likes\\
		\midrule
		\textbf{$I_{t}$}: & Index number of tweets\\
		\hline
		\bottomrule
	\end{tabular}
	\label{tab:FeaturesNotation}
\end{table}

\paragraph*{Following or Friending}
\label{Methodology:TCFF}
Following or friending are user-level features indicating that a Twitter user has subscribed to the updates of another user (i.e. following another user)~\cite{granovetter1977strength}. Following users who are not part of one's interpersonal network results in a large amount of novel information. One of the important indicators for calculating the influence score for Twitter users is the $followers/following$ ratio. The $follower/following$ ratio compares the number of $u_i$'$s$ subscribers to the number of users $u_i$ is following. Users are more interested in updates if the $follower/following$ ratio is high~\cite{anger2011measuring}. In our model, we use $friends$ as one of the indicators when assigning a social reputation score to a user. 

\paragraph*{Number of Followers}
\label{Methodology:TCNF}
Number of followers is another user-level feature that shows the number of people interested in the tweets of a specific user $u_i$. 
As discussed in~\cite{mccoy2017university}, number of followers is one of the most important parameters for measuring user influence. The more followers a Twitter user has, the more influential~\cite{leavitt2009influentials} a user is. Preussler \textit{et al.}~\cite{preussler2010managing} correlate the number of followers with the reputation of the Twitter user. According to their study, as the number of followers increases, the importance/credibility of the underlying user also increases. Based on these studies, we also considered the number of followers as an important parameter and used it as an input to calculate the social reputation of the user.

\paragraph*{Number of Retweets}
\label{Methodology:TCNR}
A tweet is considered important when it receives many positive reactions from other accounts. The reactions may be in the form of likes or retweets. Retweets act as a form of endorsement, allowing Twitter users to forward the content generated by other users, thus raising the content's visibility. Hence it is a way of promoting a topic and is associated with the reputation of the user~\cite{dutta2018retweet}.  Since retweeting is linked with popular topics and directly affects the reputation of a user, it is a key parameter for identifying possible fake account holders. As described in~\cite{gilani2017bots}, bots or fake accounts depend more on retweeting existing content rather than posting new tweets. In our model, we consider the number of retweets as one of the main parameters for assigning an influence score to user accounts. To do so, we calculate the number of times a tweet is retweeted. Additionally, we calculate the total number of tweets for each user. The total number of tweets of user $u_i$ is denoted by $N_{T}(u_i)$. Finally, we calculate the retweet ratio (using \href{https://website.grader.com/}{Twitter grader}) for each tweet by considering a tweet that is retweeted divided by the total number of tweets, given in equation~\eqref{eq:retweetratio}.

\begin{equation}
R_{R}(u_i)=\frac{Retweets\ }{N_{T}(u_i)}
\label{eq:retweetratio}
\end{equation}

\paragraph*{Likes}
\label{Methodology:TCL}
The number of likes is a reasonable proxy for evaluating the quality of a tweet. The authors in~\cite{gilani2017depth} showed that humans receive more likes per tweet when compared with bots. In~\cite{gilani2017classification}, the authors used likes as one of the metrics for classifying Twitter accounts as human or automated. As mentioned in~\cite{alrubaian2017reputation}, if a specific tweet receives a large number of likes it can be safely concluded that other users are interested in the tweets of the underlying user. Based on this observation, we calculate the liked ratio by using the number of likes for each tweet and dividing it by the total number of tweets for the corresponding user as shown in equation~\eqref{eq:likedratio}. 
\begin{equation}
l(u_i)=\frac{Liked\ tweets}{N_{T}(u_i)}
\label{eq:likedratio}
\end{equation}

\paragraph*{URLs}
\label{Methodology:TCU}
A URL is a content level feature that some users include in their tweets~\cite{hughes2009twitter}. Since tweets are limited to a maximum of~280 characters, it is common that users cannot add important information to their tweets. To overcome this issue, tweets are commonly populated with URLs pointing to a resource where more information can be found. In our model, we consider the URL as an independent variable for engagement measurements~\cite{Han}. We count the tweets that include a URL and calculate the URL ratio by considering the total number of tweets containing URLs over the total number of tweets as given in equation~\eqref{eq:urlratio}. 
\begin{equation}
U(u_i)=\frac{Tweets\ with \ URLs}{N_{T}(u_i)}
\label{eq:urlratio}
\end{equation}

\paragraph*{Listed Count}
\label{Methodology:LC}
In Twitter, a user has the option to form several groups by creating lists of different users\footnote{\url{https://help.twitter.com/en/using-twitter/twitter-lists}} (e.g. competitors, followers, colleagues, etc.). Twitter lists are mostly used to keep track of the most influential people\footnote{\url{https://www.postplanner.com/how-to-use-twitter-lists-to-always-be-engaging/}}. The simplest way to measure the influence of a Twitter user is by looking at the number of lists that the user is included on. Being a member of a large number of lists is an indicator that this user is considered important by others. Based on this assumption, in our model, we also consider the number of lists that each user belongs to.

\paragraph*{Status Counts}
\label{Methodology:STC}
Compared to the other popular OSNs, Twitter is considered to be a service that is \textit{less} social\footnote{\url{https://econsultancy.com/twitter-isn-t-very-social-study/}}. This is mainly due to the large number of inactive users or users who show low motivation in engaging in an online discussion. Twitter announced a new feature ``Status availability", that verifies the status of a user\footnote{\url{https://www.pocket-lint.com/apps/news/twitter/146714-this-is-what-twitter-s-new-online-indicators-and-status-updates-look-like}}. To this end, during the calculation of user influence scores, we also consider how active they are by measuring how often a user performs a new activity\footnote{\url{https://sysomos.com/inside-twitter/most-active-twitter-user-data/}}.

\paragraph*{Mention by Others}
\label{Methodology:MBO}
A mention within a tweet contains another person's username anywhere in the body of the tweet. 
Due to the fact that mentions indicate the inclusion of a user in conversations, tracking Twitter mentions is one of the most effective ways to measure the presence of a user in the network. The retweet and mention ratio is calculated by Isabel and Christian~\cite{anger2011measuring} to judge the reaction from other users to a user tweet. In addition, these two parameters are also used in \href{https://website.grader.com/}{Twitter Grader}~(an online tool) to assign a score to a Twitter user. In our model, we use the retweet and mention ratio along with other indicators to check how influential a Twitter user is. The mention ratio can be calculated using equation~\eqref{eq:mentionratio}.
\begin{equation}
M_{R}(u_i)=\frac{Tweets\ with\ mentions}{N_{T}(u_i)}
\label{eq:mentionratio}
\end{equation}

\paragraph*{Original Content Ratio}
\label{Methodology:OCR}
It has been observed that most Twitter users retweet posts by others~\cite{anger2011measuring} instead of posting original tweets. As a result, Twitter has become a pool of constantly updating information streams. For users with high influence
\ in the network, the best strategy is to use the \href{https://www.themuse.com/advice/20-essential-twitter-rules-youve-probably-never-heard}{$30/30/30$} rule: 30\% retweets, 30\% original content, and 30\% engagement. With this in mind, in our model we are looking for original user tweets and add them to their influence score. We calculate the original content ratio by extracting retweet posts by others from the total tweets of users as given in equation~\eqref{eq:oriconratio}.
\begin{equation}
O_{R}(u_i)=\frac{N_{T}(u_i)-Retweeted\ posts\ }{N_{T}(u_i)}
\label{eq:oriconratio}
\end{equation}

\subsection{Derived Features for Twitter Users}
\label{SS:DFTU}
In this section, we elaborate on the extraction of extra features after the consideration and evaluation of basic ones. Additionally, we discuss the sentiment analysis technique used to analyse user tweets.

By using the basic features described earlier, we calculated the following parameters for each user:

\begin{itemize}
	\item Social reputation of the user;
	\item H-index score based on likes and retweets;
	\item Sentiment score;
	\item Tweet credibility;
	\item Influence score.
\end{itemize}

\paragraph*{Social Reputation of the User}
\label{Methodology:SR}
The main factor for calculating the social reputation $R_{s}(u_i)$ of a user $u_i$ is the number of users that are interested in $u_i$'s updates. 
Hence, $u_i$'s social reputation is based on number of followers, friends and statuses~\cite{anger2011measuring, wang2010don}.

\begin{multline}
	R_{s}(u_i)=\log((1+F_{fol}(u_i)) \cdot (1+F_{fol}(u_i)))+ \\ \log(1+S(u_i))-\log((1+F(u_i))
	\label{Eq:SRU}
\end{multline}

In equation~\eqref{Eq:SRU}, $R_{s}(u_i)$ is directly proportional to $F_{fol}(u_i)$ and $S(u_i)$. Based on several studies~\cite{alrubaian2017reputation, anger2011measuring, wang2010don}, the $R_{s}(u_i)$ of a user is more dependent on $F_{fol}(u_i)$ and that is the reason we give importance to $F_{fol}(u_i)$ in comparison to $S(u_i)$ and $F(u_i)$. If a user~$(u_i)$ has a large number of followers then it is evident that the user is more reputed in his network. In addition, if a $u_i$ is more active in updating his/her status then there are more chances that the tweets from $u_i$ receive more likes and get retweeted. As $F_{fol}(u_i)$ and $S(u_i)$ increase, $u_i$'s social reputation $R_{s}(u_i)$ also increases and vice versa. Furthermore, if a user has less $F_{fol}(u_i)$ in comparison to $F(u_i)$ then obviously the $R_{s}(u_i)$ of a user is small. As can be seen from equation~\eqref{Eq:SRU}, there is an inverse relation between $R_{s}(u_i)$ and $F(u_i)$. 

\paragraph*{H-Index Score}
\label{Methodology:RI}
The h-index score is most commonly used to measure the productivity and impact of a scholar or scientist in the research community. It is based on the number of publications as well as the number of citations for each publication\footnote{\url{https://www.researchgate.net/post/How_to_calculate_h_index_for_an_author}}. In our work, we use the h-index score for a more accurate calculation of user influence scores. 
The h-index score of a Twitter user is calculated considering the number of likes and retweets for each tweet. To find the h-index score\footnote{\url{https://gallery.azure.ai/Notebook/Computing-Influence-Score-for-Twitter-Users-1}}, we sort the tweets based on the number of likes and retweets (in decreasing order). 

Algorithm~\ref{alg:Rhindex} describes the main steps for calculating the h-index score of a Twitter user based on the \textit{number of retweets} and \textit{likes}.

\begin{algorithm}
	\caption{Calculating a $User_i$ h-index score for retweets and likes}\label{alg:Rhindex}
	\begin{algorithmic}[1]
		\Procedure{H-Index score}{$hindex(u_i)$}
		\State Arrange $R_{n}$/$L_{n}$ for each tweet of the user in decreasing order 
		\For{\texttt{$I_{t}$ in list:}}
		\If{$R_{n}$/$L_{n}$ of a tweet $<$ $I_{t}$}
		\State return $I_{t}$
		\EndIf
		\EndFor
		\State return number of tweets
		\EndProcedure
	\end{algorithmic}
\end{algorithm}
$R_{hindex}(u_i)$ and $L_{hindex}(u_i)$ are novel features used to measure the relative importance of a user on Twitter. A tweet that has been retweeted many times and liked by many users is considered to be attractive to readers~\cite{alrubaian2017reputation, riquelme2016measuring}. For this reason, we use $R_{hindex}(u_i)$ and $L_{hindex}(u_i)$ to measure the influence of a Twitter user. The higher the $R_{hindex}(u_i)$ and $L_{hindex}(u_i)$ score of a Twitter user, the higher is that user's influence. 

\paragraph*{Credibility of Twitter Users}
\label{Methodology:credibiltytwitteruser}
Credibility refers to believability~\cite{castillo2011information}, which requires reasonable grounds for being believed. Twitter user credibility can be assessed by using the information available on the Twitter platform. In our research work, we use both the sentiment score and tweet credibility in identifying credible Twitter users. 

\smallskip

\underline{Sentiment Score}:
\label{Methodology:CNS}
It has been observed that OSNs are a breeding ground for the distribution of fake news where even individual Twitter posts can have a significant impact~\cite{wolfsfeld2013social} that will affect the outcome of an event. 

With this in mind, we used \href{https://www.earthdatascience.org/courses/earth-analytics-python/using-apis-natural-language-processing-twitter/analyze-tweet-sentiments-in-python/}{sentiment analysis} and the TextBlob~\cite{loria2014textblob} library  to analyse recent tweets with the main aim of identifying certain attitudes/patterns that can lead us to the identification of credible news.
The sentiment analysis returned a score using polarity values ranging from 1 to -1 and helps in tweet classification. We classify the collected tweets as \textit{(1)} Positive~$P_{+ve}$ \textit{(2)} Neutral~$N_{+ve}$, and \textit{(3)} Negative~$N_{-ve}$ based on the number of positive, neutral and negative words in a tweet with $P_{+ve}(u_i)$ being the most credible tweets followed by the neutral tweets $N_{+ve}(u_i)$ and then the least credible tweets $N_{-ve}(u_i)$. According to Morozov~\textit{et al.}~\cite{morozov2014analysing}, the least credible tweets are associated with negative social events. They have more negative words and opinions, while credible tweets have more positive opinions and words.

After classification, based on previous tweets, we assign a sentiment score to each user~$(u_i)$~\cite{alrubaian2017reputation} using the following equation:  

\begin{equation}
C_{s}(u_i)=\frac{\sum N_{+ve}+ \sum P_{+ve}}{\sum N_{+ve}+ \sum P_{+ve} + \sum N_{-ve}}
\label{eq:NT}
\end{equation}

\smallskip

\underline{Tweet Credibility}:
\label{Methodology:CTS}
Donovan~\cite{odonovan2012credibility} focused on finding the best indicators for credibility. According to these results, the best indicators for tweet credibility are URLs, mentions, retweets and length. Gupta \textit{et al.}~\cite{gupta2012credibility} ranked tweets based on tweet credibility. The parameters used as input for the ranking algorithm are total unique users, tweets, tweets with URLs, single tweets, retweets, trending topics, start and end date. Based on the existing literature, we compute the credibility of tweets by considering $R_{R}(u_i)$, $L_{R}(u_i)$, $H_{R}(u_i)$, $U(u_i)$ and $O_R(u_i)$:

%

Based on the above parameters, we measure tweet credibility by using (equation~\eqref{eq:TC}).

\begin{equation}
\begin{aligned}
\resizebox{1\hsize}{!}{$Twt_{cr}(u_i)=\frac{((R_{R}(u_i)+L_R(u_i)+H(u_i)+ U(u_i))}{4} \cdot O_R(u_i))$}
\end{aligned}
\label{eq:TC}
\end{equation}

For calculating the credibility of tweets, first we extract the $O_R(u_i)$ published by a Twitter user as a tweet. Then we consider the number of times this tweet is $R_{R}(u_i)$ and $L_R(u_i)$ by others. In addition, we also consider $H(u_i)$ and $U(u_i)$ as they are the functions that can be used for user engagement. Since these four parameters, $R_{R}(u_i)$, $L_R(u_i)$, $H(u_i)$ and $U(u_i)$, are linked with  $O_R(u_i)$, we start by calculating the average of these four parameters and then multiply that result by $O_R(u_i)$. Based on these parameters, we calculate the credibility of tweets as given in equation~\eqref{eq:TC}.
 
\subsection{Influence Score}
\label{Methodology:RS}
The influence score of a Twitter user is calculated based on the evaluation of \textit{both} content and context features. More precisely, we consider the following parameters described earlier: 
$R_{s}(u_i)$, $C_{s}(u_i)$, $Twt_{cr}(u_i)$ and $hindex(u_i)$. 
After calculating the values of all of these features, we use them as input for Algorithm~\ref{alg:reputation}~line~7. which calculates the influence score for the underlying Twitter user. 

\textit{Equation Formulation:}
\label{SSC:EF}
To find out how influential a Twitter user is, researchers have taken into consideration one, two or more of the following characteristics:

\begin{itemize}
	\item Weight-age assigned to their tweets and impact~\cite{alrubaian2017reputation};
	\item Credibility of the tweets~\cite{odonovan2012credibility, alrubaian2017reputation};
	\item Social reputation of the Twitter user~\cite{garcia2017understanding};
	\item Level of activity, involvement in follow-up and discussions and the ability to propose new ideas~\cite{chen2011tweet}.
\end{itemize}

An influential Twitter user must be highly active (e.g. able to start new discussions, have ideas that impact other users' behaviors, etc.). Additionally, the user's tweets must be credible, relevant and highly influential (i.e. liked and retweeted by a large number of other users). If the tweets of highly influential users are credible and the polarity of their tweet content is positive, they are highly acknowledged and recognised by the community. In short, for a Twitter user to be considered influential, we combine the efforts of~\cite{alrubaian2017reputation, odonovan2012credibility, garcia2017understanding, chen2011tweet} and calculate the influence score through Algorithm~\ref{alg:reputation}~line~7.

Algorithm~\ref{alg:reputation} shows the steps we follow to calculate the influence score of a user. 

\begin{algorithm}
	\caption{Calculating the User $u_{i}$ Influence Score}\label{alg:reputation}
	\begin{algorithmic}[1]
		
		\Procedure{Influence Score} {$R_{s}(u_i)$}\medskip
		\State For $i^{th}$ user $u_i$ \medskip
		\State Calculate $R_{hindex}$ and $L_{hindex}$ of User $u_{i}$ by using Algorithm~\ref{alg:Rhindex} \medskip
		\State Calculate Sentiment Score of User $u_{i}$ \smallskip
		\begin{center}
			$C_{s}(u_i)=\frac{\sum N_{+ve}+ \sum P_{+ve}}{\sum N_{+ve}+ \sum P_{+ve} + \sum N_{-ve}}$
		\end{center}\medskip
		\State Calculate Tweet Credibility of User $u_{i}$ \smallskip
		\begin{center}
			$Twt_{cr}(u_i)=[\frac{R_{R}(u_i)+L_R(u_i)+H_R(u_i)+U_{R}(u_i)}{4}] \cdot O_R(u_i))$ 
		\end{center} \medskip
		\State Calculate Social Reputation of User $u_{i}$ \smallskip
		\begin{center}
			$R_{s}(u_i)=\log((1+F_{fol}(u_i)) \cdot (1+F_{fol}(u_i))) + \log(1+S(u_i))-\log((1+F(u_i))$
		\end{center} \medskip
		\State Compute Influence Score of User $u_{i}$ \smallskip
		\begin{center}
			$R(u_i)=\frac{C_{s}(u_i)+Twt_{cr}(u_i)+R_{s}(u_i)+R_{hindex}(u_i)+L_{hindex}(u_i)}{5}$
		\end{center} \medskip
		\EndProcedure
	\end{algorithmic}
\end{algorithm}

\subsection{Parameter Selection and Comparison with Previous Models}
\label{SC:PSCM}
The parameters used for calculating the influence score are based on an extensive study of the existing literature. The selected parameters are used for detection purposes~\cite{fazil2018hybrid, amleshwaram2013cats, yang2013empirical}, assigning a score~\cite{gupta2014tweetcred} or for classification purposes~\cite{gilani2017depth}. We used all these parameters to assign an influence score to users. Table~\ref{tab:Comp table} provides an overview of comparisons between existing models based on feature selection.

\begin{table*}[!ht]
	\tiny
	\caption{{Comparison of Models using Feature Selection}}
	\label{tab:Comp table}
	\begin{tabular}{|l|p{.8cm}|p{.45cm}|p{.45cm}|p{1.25cm}|p{1.25cm}|p{.55cm}|p{.65cm}|p{.65cm}|p{.5cm}|p{.5cm}|p{.65cm}|p{.65cm}|p{.68cm}|p{.53cm}|p{.63cm}|p{.45cm}|}
		\hline
		Papers & \multicolumn{3}{l|}{$R_{s}(u_i)$}    & \multicolumn{2}{l|}{$h_Index$}        & $C_{s}(u_i)$ & \multicolumn{6}{l|}{$Twt_{cr}(u_i)$}                                            & \multicolumn{4}{l|}{URLs, List and Mentions}     \\ \hline
		& $F_{fol}(u_i)$ & $F(u_i)$ & $S(u_i)$ & $R_{hindex}(u_i)$ & $L_{hindex}(u_i)$ &              & $R_{R}(u_i)$ & $L_{R}(u_i)$ & $H(u_i)$ & $U(u_i)$ & $O_{R}(u_i)$ & $N_{T}(u_i)$ & $M_{R}(u_i)$ & $M(u_i)$ & $U_{R}(u_i)$ & $L(u_i)$ \\ \hline
		
		\cite{alrubaian2017reputation}     &   \checkmark  & \checkmark & &  &   &   \checkmark&  \checkmark   &    &&  &  & \checkmark  &  & \checkmark& \checkmark &\\ \hline
		
		\cite{gupta2014tweetcred}  &   \checkmark  & \checkmark && &&  \checkmark &  \checkmark   &   \checkmark  &\checkmark&  \checkmark&  \checkmark&  &  \checkmark & \checkmark& \checkmark &\\ \hline
		
		\cite{gilani2017depth}&   \checkmark  & \checkmark & &  &   &  \checkmark &  \checkmark   &   \checkmark  &&  &  \checkmark&  \checkmark &   & \checkmark& \checkmark &\\ \hline
		
		\cite{amleshwaram2013cats}      &   \checkmark   &\checkmark&          &                   &                   &  \checkmark &              &              &          &          &  \checkmark&              & \checkmark & \checkmark &  \checkmark &          \\ \hline
		\cite{yang2013empirical}     &   \checkmark  & \checkmark &          &                   &                   &  \checkmark  & \checkmark  &              &  \checkmark & \checkmark  &              &  \checkmark  &              &          &  \checkmark   &          \\ \hline
		\cite{fazil2018hybrid}      & \checkmark  &  \checkmark &          &                   &                   &    \checkmark&   \checkmark&              &  \checkmark&  \checkmark &              &              &  \checkmark  &  \checkmark& \checkmark &          \\ \hline
		Proposed     &   \checkmark  & \checkmark &\checkmark & \checkmark &  \checkmark &  \checkmark &  \checkmark   &   \checkmark  &\checkmark&  \checkmark&  \checkmark&  \checkmark &  \checkmark & \checkmark& \checkmark &\checkmark\\ \hline
	\end{tabular}
\end{table*}


\section{Active learning and ML Models}
\label{sec:ActiveLearning}
In the existing literature, the classification of Twitter users is primarily performed on a manually annotated dataset. A manually annotated dataset gives ground truth, however, manual labeling is an expensive and time-consuming task. In our proposed approach, we used active learning, a semi-supervised ML model that helps in classification when the amount of available labeled data is small. In this model, the classifier is trained with a small amount of training data (labeled data points). Then the points ambiguous to the classifier in the large pool of unlabeled data points are labeled and added to the training set~\cite{Settles}. This process is repeated until all the ambiguous instances are queried or the model performance does not improve above a certain threshold. Based on the proposed model, we train our classifier on a small human-annotated dataset which further classifies a large pool of unlabeled data points efficiently and accurately.

Our active learning process evolves through the following steps:

\begin{itemize}
    \item \textbf{Data Gathering:} First, we gather the unlabeled data for 50,000 Twitter users. The unlabeled data is split into a seed -- a small labeled dataset~(manually labeled) and a large pool of unlabeled data. The seed is used to train the classifier just like a normal ML model. Using a dataset~(seed) of 1,000 manually annotated data, we classify each political Twitter user as a trustworthy or untrustworthy user. 
    \item \textbf{Selection of Unlabeled Instances:} A pool-based sampling with a batch size of 100 is used in which 100 ambiguous instances from the unlabeled dataset are labeled and added to a labeled dataset. Different sampling is employed to select the instances from the unlabeled dataset. For the new labeled dataset, the classifier is re-trained and then the next batch of ambiguous unlabeled instances for labeling are selected. The process is repeated until the model performance does not improve above a certain threshold.
\end{itemize}

\medskip 

In addition, we used the following two classifiers:
\begin{itemize}
	\item \textbf{Random Forest Classifier (RFC):} RFC is an ensemble tree-based learning algorithm~\cite{Biau}. It aggregates the votes from different decision trees to decide the output class of the instance. RFC can run efficiently on large datasets, handle thousands of input variables, measure the relative importance of each feature, and produces a highly accurate classifier.
	\item \textbf{Support Vector Machine (SVM):} SVM produces high accuracy with less computation power and is widely used in classification tasks. To classify the instances, the SVM finds a hyperplane in $N$-dimensional space, where $N$ represents the number of features~\cite{Noble}. The goal of SVM is to perform classification by finding the hyperplane separating the two classes more accurately (maximising the margin between two classes).
\end{itemize}

\section{Experimental Results and Model Evaluation}
\label{sec:Evaluation}

\noindent \textbf{Experimental Setup:}
To extract the features from Twitter and generate the dataset we used Python~3.5. The python script was executed locally on a machine with the following configuration:
Intel Core i7, 2.80*8 GHZ,~32GB, Ubuntu~16.04 LTS~64~bit.
For the training and evaluation of the machine learning models, we switched to \href{https://colab.research.google.com/notebooks/welcome.ipynb\#recent=true}{Google Colab}. We use the modAL framework~\cite{Danka}, which is an active learning framework for python. It is a flexible, modular, and extensible framework built on top of Scikit learn. For the learner to query the instance labels, we use pool-based sampling and for the query strategy, we use different sampling techniques. For classification purposes, we use two RFC and SVC classifiers implemented using the scikit-learn library.

\subsection{Dataset and Data Collection}
\label{Eval:DC}  
To collect user features and tweets we used \href{https://pypi.org/project/tweepy/}{tweepy} -- Twitter's search API. Tweepy has certain limitations, as it only allows the collection of a certain number of parameters. Additionally, there is also a data rate limit that prevents the collection of information above a certain threshold. In our dataset, we chose to analyse the Twitter accounts of~50,000 politicians. 

The main reason we decided to evaluate the profiles of politicians is their intrinsic potential to influence public opinion as their content originates and exists in a sphere of political life which is, unfortunately, often surrounded by controversial events and outcomes. 
When selecting these politicians, we only considered those with a \textit{public profile} while users that seemed to be \textit{inactive} (e.g. limited number of followers and activities) were omitted. Finally, for each user we extracted all the necessary features required by our model. 

Using the extracted features and tweets, we calculated an influence score for each user. Furthermore, we generated a dataset consisting of~19 features including the influence score for~50,000 Twitter users. There are features such as the number of followers, likes, etc. which have no defined upper limits and may have outlying values. Hence, for these features, we used a percentile clip. We then normalised our features using min-max normalisation, with~0 being the smallest and~1 being the largest value. 

\subsection{Performance Measurements of Machine Learning and Neural Network Models}
\label{PMML}
We garnered 50,000 unlabeled instances of Twitter users. The dataset was divided into three sets: training, testing, and unlabeled pool data. For the training and testing cohorts, we had 1,000 data points that were manually annotated. The rest of the data was unlabeled (49,000 instances). For the classification, we used the RFC and SVM classifiers (both classifiers are trained on the labeled dataset). Accuracy (\%) is used as the evaluation metric for model performance, which measures the percentage of correctly classified instances. To improve model accuracy, the active learner randomly selects ambiguous data points from the unlabeled data pool using three different sampling techniques and a person manually annotates the selected data. The annotated data is then added to the labeled dataset. This process is repeated 100 times for both the classifiers. The respective sampling techniques and accuracy obtained for both classifiers are discussed below.\\

\begin{figure*}[!ht]
	\begin{subfigure}{.4\textwidth}
		\centering
		\includegraphics[width=.5\linewidth]{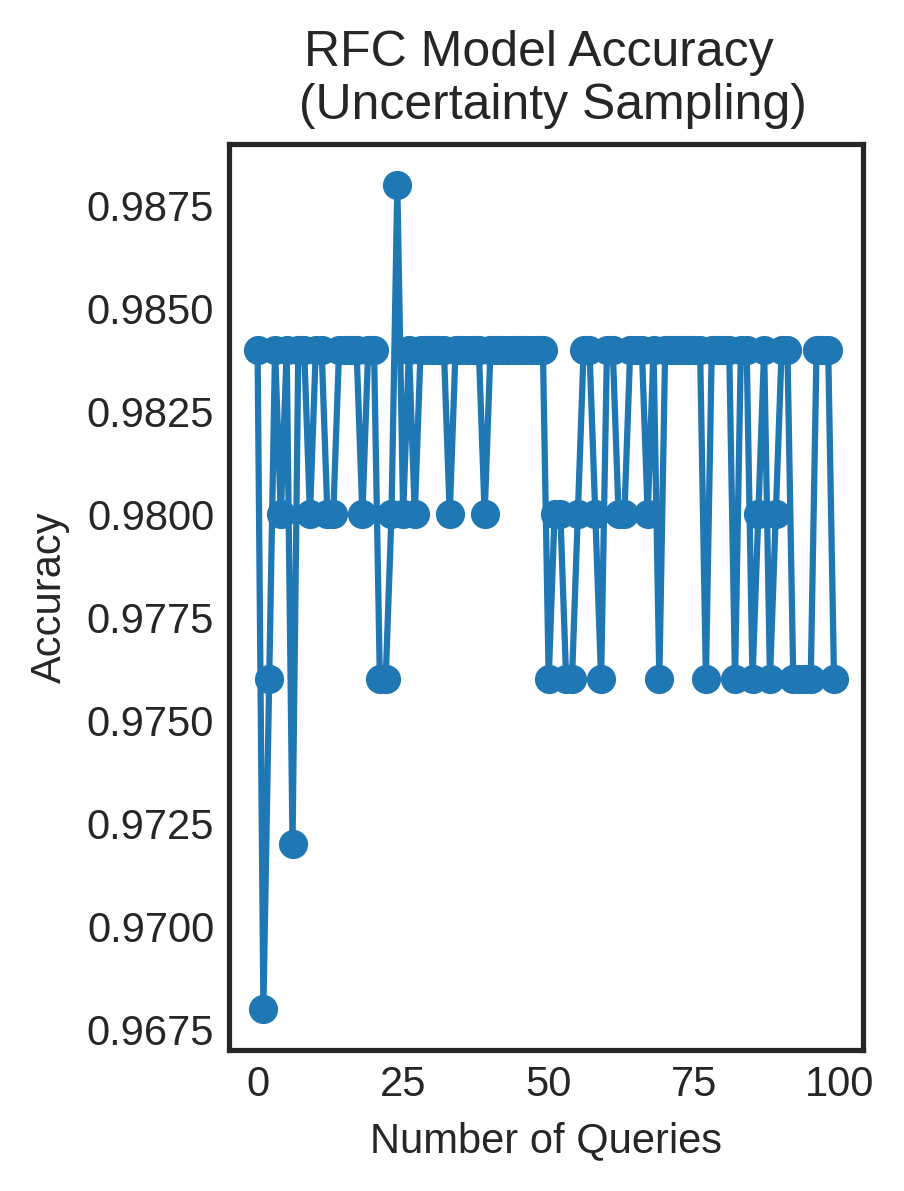}
		\caption{RFC Model Accuracy Using Uncertainty Sampling}\label{fig:RFCUP}
	\end{subfigure}
	\begin{subfigure}{.4\textwidth}
		\centering
		\includegraphics[width=.5\linewidth]{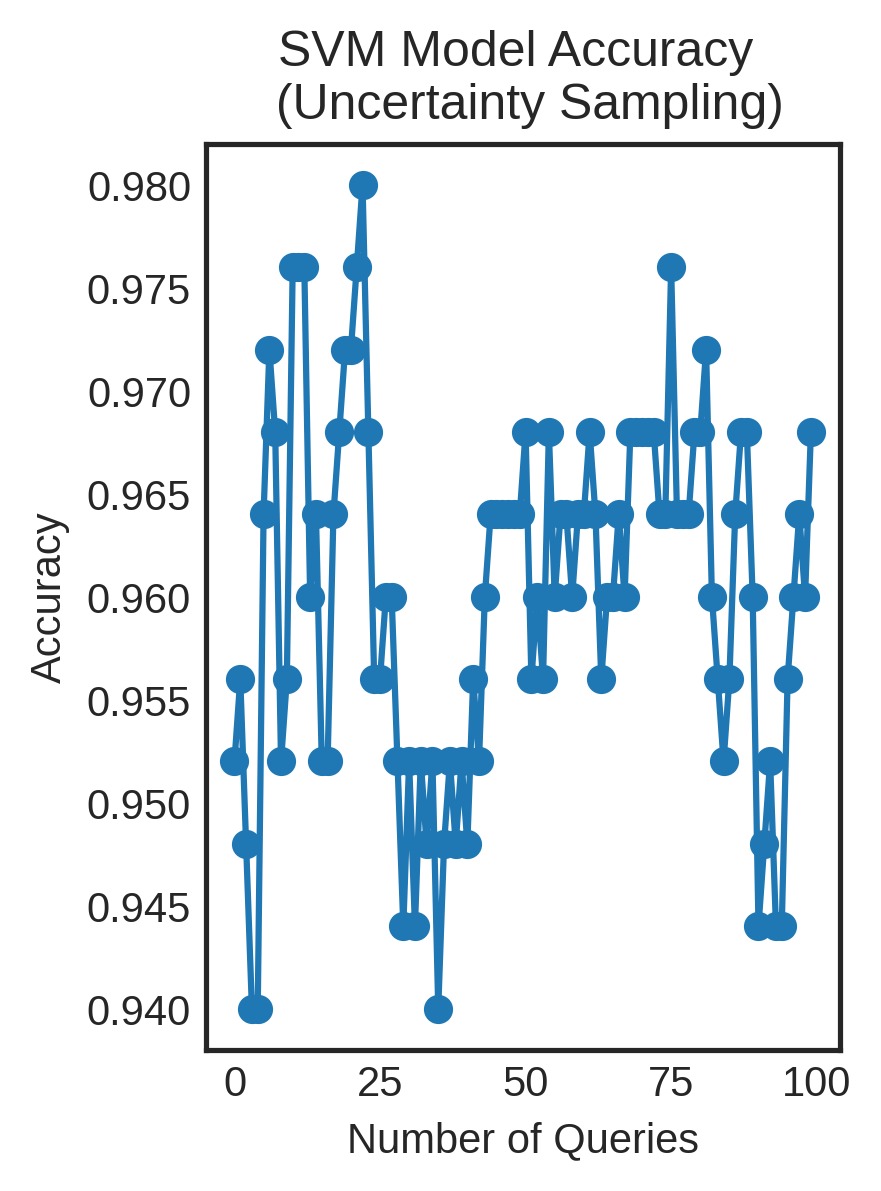}
		\caption{SVM Model Accuracy Using Uncertainty Sampling}\label{fig:SVM}
	\end{subfigure}
	\caption{Uncertainty Sampling}
\end{figure*}

\paragraph*{Uncertainty Sampling}
In uncertainty sampling, the instance in which there is the least confidence is most likely to be considered. In this type of sampling method, the most probable labels are considered and the rest are discarded. The results are shown in figure~\ref{fig:RFCUP} for the RFC, which achieves an accuracy of 97.6\% while the SVM obtained an accuracy of 96.8\% (this is shown in figure~\ref{fig:SVM}).

%
%

\begin{figure*}[!ht]
	\begin{subfigure}{.4\textwidth}
		\centering
		\includegraphics[width=.5\linewidth]{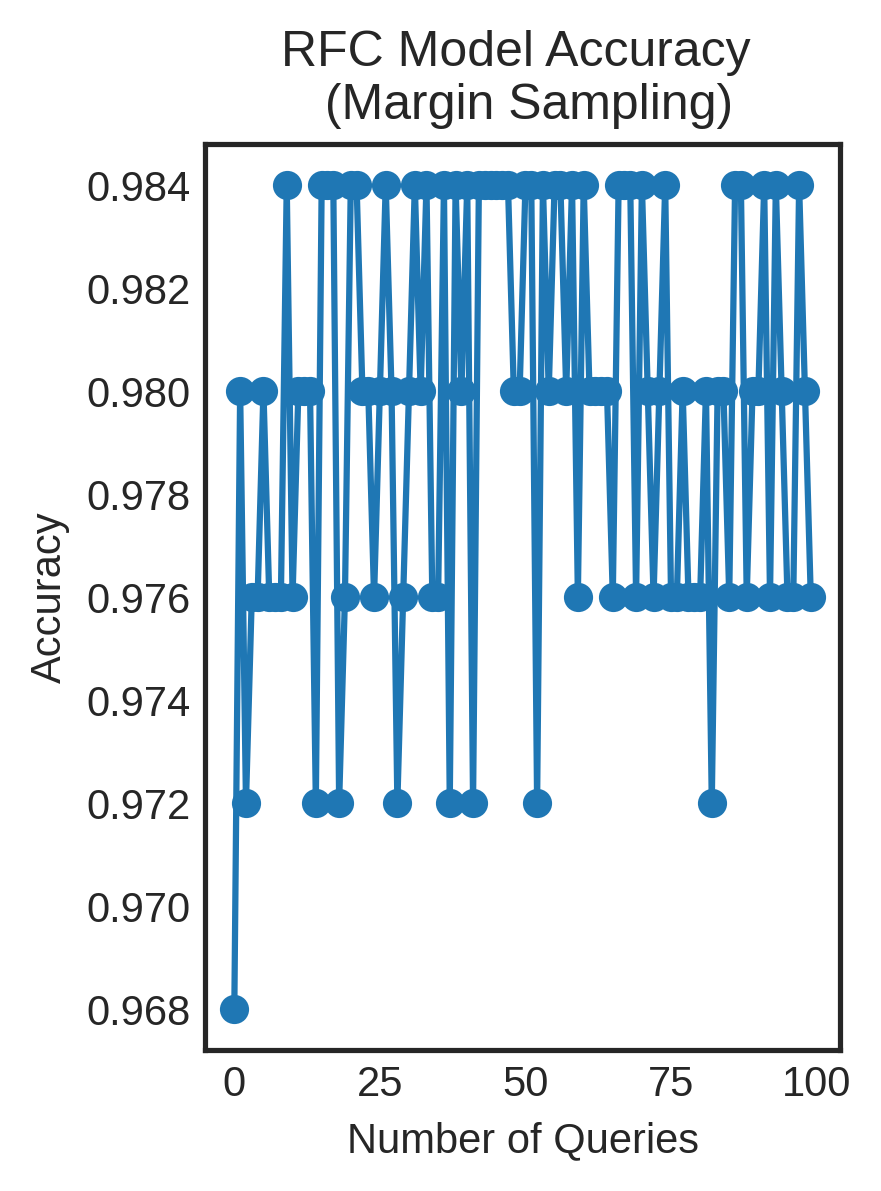}
		\caption{RFC Model Accuracy Using Margin Sampling}\label{fig:RFCMS}
	\end{subfigure}
	\begin{subfigure}{.4\textwidth}
		\centering
		\includegraphics[width=.5\linewidth]{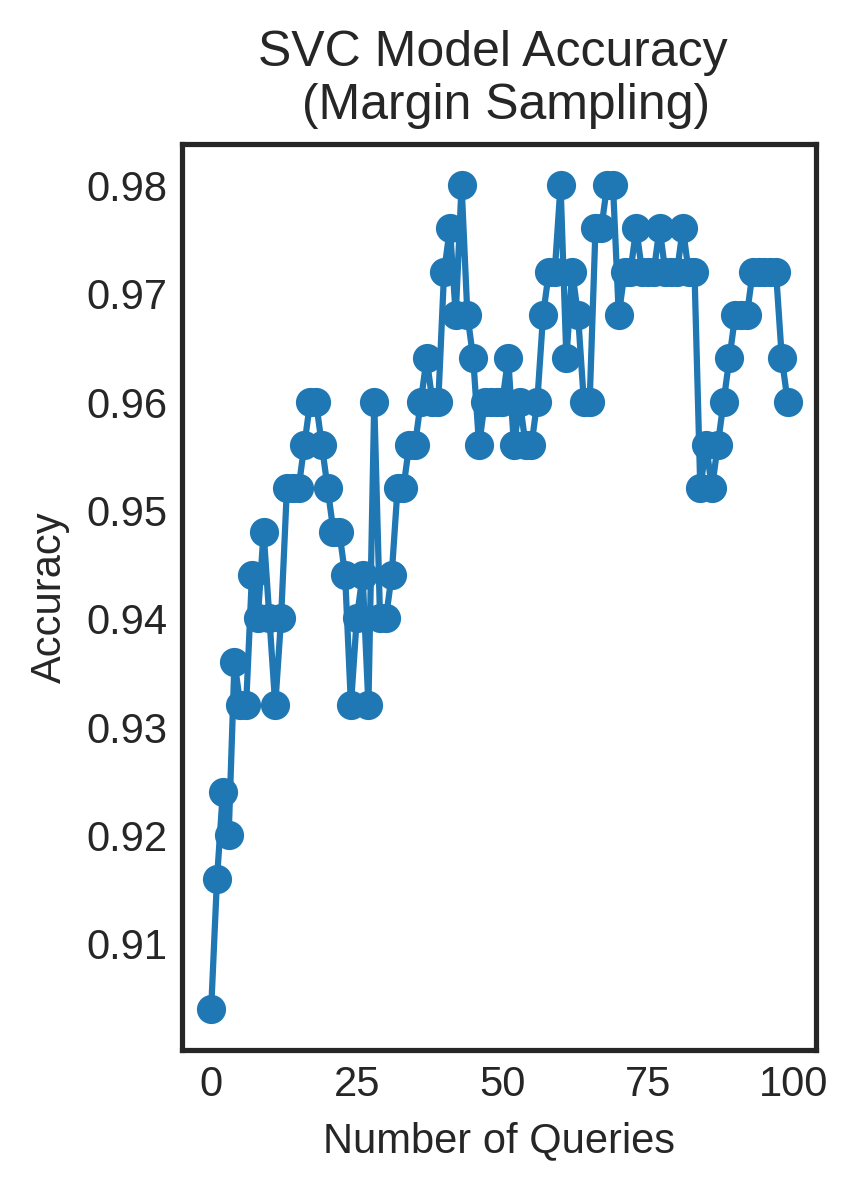}
		\caption{SVM Model Accuracy Using Margin Sampling}\label{fig:SVCMS}
	\end{subfigure}
	\caption{Margin Sampling}
\end{figure*}

\paragraph*{Margin Sampling}
In margin sampling, instances with the smallest difference between the first and second most probable labels are considered. The accuracy for RFC and SVM using margin sampling is 97.6\% and 96.2\%, respectively. This can be seen in figure~\ref{fig:RFCMS} for RFC and figure~\ref{fig:SVCMS}.
%
%

\begin{figure*}[!ht]
	\begin{subfigure}{.4\textwidth}
		\centering
		\includegraphics[width=.5\linewidth]{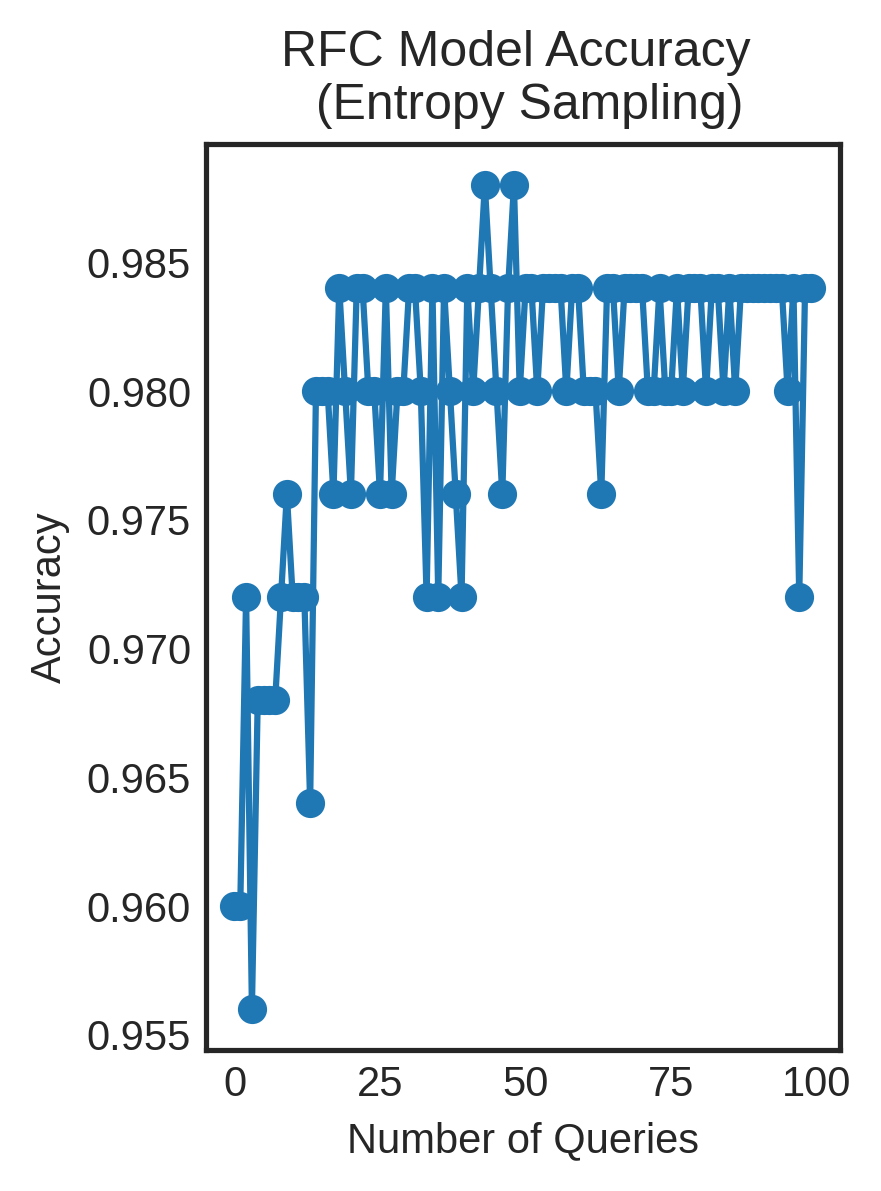}
		\caption{RFC Model Accuracy Using Entropy Sampling}\label{fig:RFCES}
	\end{subfigure}
	\begin{subfigure}{.4\textwidth}
		\centering
		\includegraphics[width=.5\linewidth]{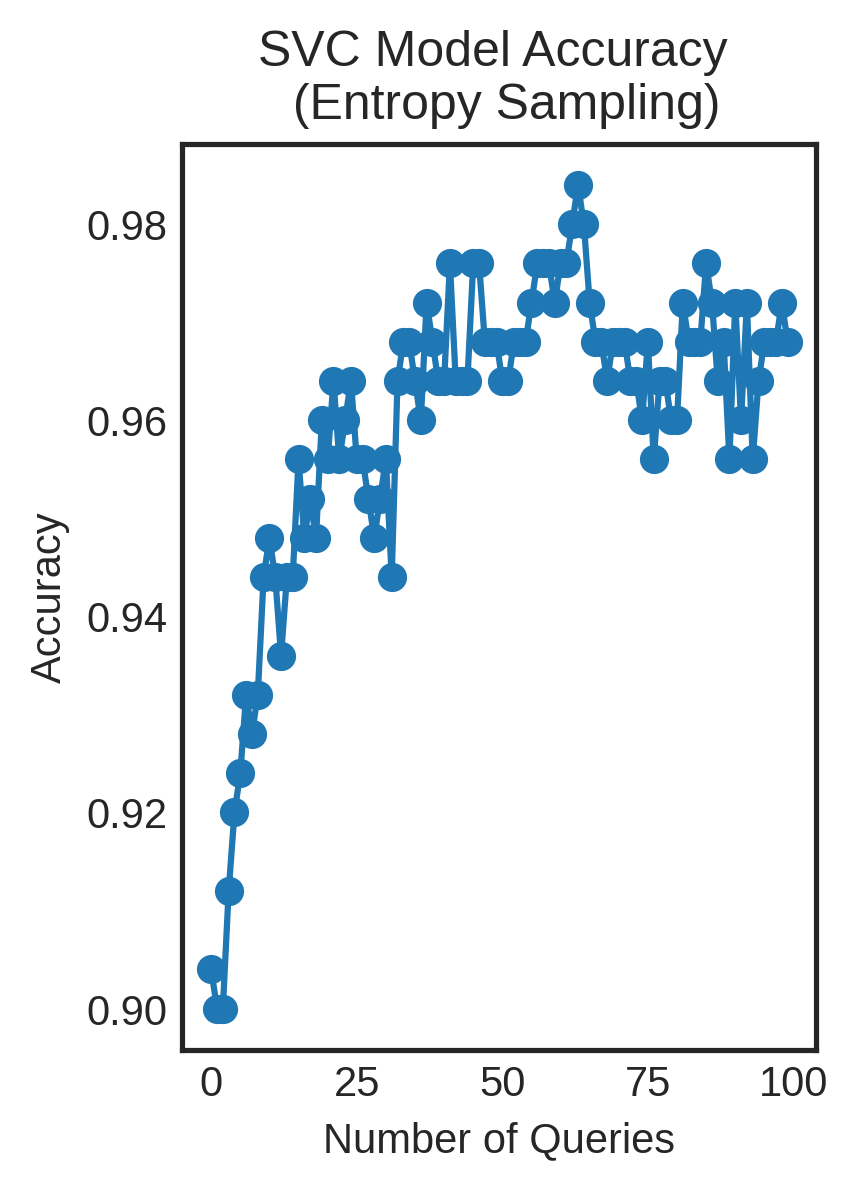}
		\caption{SVM Model Accuracy Using Entropy Sampling}\label{fig:SVCES}
	\end{subfigure}
	\caption{Entropy Sampling}
\end{figure*}

\paragraph*{Entropy Sampling}
Lastly, the entropy sampling method offers the best results, obtaining an accuracy of 98.4\% for RFC and 97\% for SVM. An explanation for this improved performance can be attributed to the fact that entropy sampling utilises all possible label probabilities, unlike the other sampling methods. For RFC and SVM, this is also shown in figure \ref{fig:RFCES} and \ref{fig:SVCES}.

%
%

\paragraph*{\textbf{Open Science \& Reproducible Research}} 
As a way to support open science and reproducible research and give the opportunity to other researchers to use, test and hopefully extend/enhance our models, we plan to make both our datasets as well as the code of our models available through the \href{https://zenodo.org/record/7014109}{Zenodo} research artifacts portal. This will not violate \href{https://developer.twitter.com/en.html}{Twitter's developer terms}. 
However, in order to maintain our anonymity, we will make this available in a camera-ready version if the paper is accepted.

\section{Conclusion}
\label{sec:Conclusion}
Having identified the significant impact of fake news on our lives, this work focused on finding ways to identify this kind of information and notify users about the possibility that a specific post from a Twitter user may not be credible. To do so, we designed a model that analyses Twitter users and assigns each a calculated score based on their social profiles, tweet credibility, sentiment score, and h-index score (retweets and likes). Users with a higher score are not only considered more influential but their tweets are also considered to have greater credibility. 
To achieve our goal, we generated a dataset of~50,000 Twitter users (politicians) along with a set of~19 features for each user. Then, we classified each Twitter user as trustworthy or untrustworthy using RFC and SVM classifiers. Moreover, we employed the active learner approach to label ambiguous unlabelled data points. During the evaluation of our models, we conducted extensive experiments using three sampling methods which show the effectiveness of our approach. We believe this work is an important step towards re-establishing user trust in social networks and a stepping stone towards building new bonds of trust between users.


We see this work as an important step towards engendering user trust in social networks and we believe that it can constitute the underpinnings for establishing trust relationships between users.

\balance


\begin{thebibliography}{00}

\bibitem{hindman2018disinformation} M. Hindman and V. Barash, “Disinformation, and Influence Campaigns on Twitter,” 2018.
	
\bibitem{java2007we} A. Java, X. Song, T. Finin, and B. Tseng, “Why we twitter: understanding microblogging usage and communities,” in Proceedings of the 9th WebKDD and 1st SNA-KDD 2007 workshop on Web mining and social network analysis, 2007, pp. 56–65.

\bibitem{adamic2005political} L. A. Adamic and N. Glance, “The political blogosphere and the 2004 US election: divided they blog,” in Proceedings of the 3rd international workshop on Link discovery, 2005, pp. 36–43.

\bibitem{ratkiewicz2011detecting} J. Ratkiewicz, M. D. Conover, M. Meiss, B. Gonçalves, A. Flammini, and F. M. Menczer, “Detecting and tracking political abuse in social media,” in Fifth international AAAI conference on weblogs and social media, 2011.

\bibitem{Michalas:14:StR} Tassos Dimitriou and Antonis Michalas. “Multi-Party Trust Computation in Decentralized Environments”. Proceedings of the 5th IFIP International Conference on New Technologies, Mobility \& Security (NTMS’12), Istanbul, Turkey, 2012.

\bibitem{Michalas:14:StRM} Tassos Dimitriou and Antonis Michalas. “Multi-Party Trust Computation in Decentralized Environments in the Presence of Malicious Adversaries”. Ad Hoc Networks Journal, a special issue on “Smart Solutions for Mobility Supported Distributed and Embedded Systems”, Elsevier, 2014.

\bibitem{wang2010don} A. H. Wang, “Don’t follow me: Spam detection in twitter,” in 2010 international conference on security and cryptography (SECRYPT), 2010, pp. 1–10.

\bibitem{jain2015hashjacker} N. Jain, P. Agarwal, and J. Pruthi, “HashJacker-detection and analysis of hashtag hijacking on Twitter,” Int. J. Comput. Appl., vol. 114, no. 19, 2015.

\bibitem{grier2010spam} C. Grier, K. Thomas, V. Paxson, and M. Zhang, “@ spam: the underground on 140 characters or less,” in Proceedings of the 17th ACM conference on Computer and communications security, 2010, pp. 27–37.

\bibitem{allcott2017social} H. Allcott and M. Gentzkow, “Social media and fake news in the 2016 election,” J. Econ. Perspect., vol. 31, no. 2, pp. 211–236, 2017.

\bibitem{metzgar2009social} E. Metzgar and A. Maruggi, “Social media and the 2008 US presidential election.,” J. New Commun. Res., vol. 4, no. 1, 2009.

\bibitem{saaya2019development} Z. Saaya and T. W. Hong, “THE DEVELOPMENT OF TRUST MATRIX FOR RECOGNIZING RELIABLE CONTENT IN SOCIAL MEDIA,” Int. J. Comput., vol. 18, no. 1, pp. 60–66, 2019.

\bibitem{zhou2019fake} X. Zhou, A. Jain, V. V Phoha, and R. Zafarani, “Fake News Early Detection: A Theory-driven Model,” arXiv Prepr. arXiv1904.11679, 2019.

\bibitem{tschiatschek2018fake} S. Tschiatschek, A. Singla, M. Gomez Rodriguez, A. Merchant, and A. Krause, “Fake news detection in social networks via crowd signals,” in Companion Proceedings of the The Web Conference 2018, 2018, pp. 517–524.

\bibitem{bovet2019influence} A. Bovet and H. A. Makse, “Influence of fake news in Twitter during the 2016 US presidential election,” Nat. Commun., vol. 10, no. 1, p. 7, 2019.

\bibitem{al2015new} M. Al-Qurishi, R. Aldrees, M. AlRubaian, M. Al-Rakhami, S. M. M. Rahman, and A. Alamri, “A new model for classifying social media users according to their behaviors,” in 2015 2nd World Symposium on Web Applications and Networking (WSWAN), 2015, pp. 1–5.

\bibitem{canini2011finding} K. R. Canini, B. Suh, and P. L. Pirolli, “Finding credible information sources in social networks based on content and social structure,” in 2011 IEEE Third International Conference on Privacy, Security, Risk and Trust and 2011 IEEE Third International Conference on Social Computing, 2011, pp. 1–8.

\bibitem{gupta2012evaluating} M. Gupta, P. Zhao, and J. Han, “Evaluating event credibility on twitter,” in Proceedings of the 2012 SIAM International Conference on Data Mining, 2012, pp. 153–164.

\bibitem{riquelme2016measuring} F. Riquelme and P. González-Cantergiani, “Measuring user influence on Twitter: A survey,” Inf. Process. \& Manag., vol. 52, no. 5, pp. 949–975, 2016.

\bibitem{liu2014tweets} Y. Liu, C. Kliman-Silver, and A. Mislove, “The tweets they are a-changin’: Evolution of Twitter users and behavior,” in Eighth International AAAI Conference on Weblogs and Social Media, 2014.

\bibitem{tinati2012identifying} R. Tinati, L. Carr, W. Hall, and J. Bentwood, “Identifying communicator roles in twitter,” in Proceedings of the 21st International Conference on World Wide Web, 2012, pp. 1161–1168.

\bibitem{moens2014mining} M.-F. Moens, J. Li, and T.-S. Chua, Mining user generated content. Chapman and Hall/CRC, 2014.

\bibitem{rao2010classifying} D. Rao, D. Yarowsky, A. Shreevats, and M. Gupta, “Classifying latent user attributes in twitter,” in Proceedings of the 2nd international workshop on Search and mining user-generated contents, 2010, pp. 37–44.

\bibitem{uddin2014understanding} M. M. Uddin, M. Imran, and H. Sajjad, “Understanding types of users on Twitter,” arXiv Prepr. arXiv1406.1335, 2014.

\bibitem{al2011experimental} H. S. Al-Khalifa and R. M. Al-Eidan, “An experimental system for measuring the credibility of news content in Twitter,” Int. J. Web Inf. Syst., vol. 7, no. 2, pp. 130–151, 2011.

\bibitem{granovetter1977strength} M. S. Granovetter, “The strength of weak ties,” in Social networks, Elsevier, 1977, pp. 347–367.

\bibitem{anger2011measuring} I. Anger and C. Kittl, “Measuring influence on Twitter,” in Proceedings of the 11th International Conference on Knowledge Management and Knowledge Technologies, 2011, p. 31.

\bibitem{dutta2018retweet} H. S. Dutta, A. Chetan, B. Joshi, and T. Chakraborty, “Retweet us, we will retweet you: Spotting collusive retweeters involved in blackmarket services,” in 2018 IEEE/ACM International Conference on Advances in Social Networks Analysis and Mining (ASONAM), 2018, pp. 242–249.

\bibitem{grandjean2016social} M. Grandjean, “A social network analysis of Twitter: Mapping the digital humanities community,” Cogent Arts \& Humanit., vol. 3, no. 1, p. 1171458, 2016.

\bibitem{shu2017fake} K. Shu, A. Sliva, S. Wang, J. Tang, and H. Liu, “Fake news detection on social media: A data mining perspective,” ACM SIGKDD Explor. Newsl., vol. 19, no. 1, pp. 22–36, 2017.

\bibitem{gupta2014tweetcred} A. Gupta, P. Kumaraguru, C. Castillo, and P. Meier, “Tweetcred: Real-time credibility assessment of content on twitter,” in International Conference on Social Informatics, 2014, pp. 228–243.

\bibitem{mendoza2010twitter} M. Mendoza, B. Poblete, and C. Castillo, “Twitter under crisis: Can we trust what we RT?,” in Proceedings of the first workshop on social media analytics, 2010, pp. 71–79.

\bibitem{gupta20131} A. Gupta, H. Lamba, and P. Kumaraguru, “1.00 per rt bostonmarathon prayforboston: Analyzing fake content on twitter,” in 2013 APWG eCrime researchers summit, 2013, pp. 1–12.

\bibitem{gupta2013faking} A. Gupta, H. Lamba, P. Kumaraguru, and A. Joshi, “Faking sandy: characterizing and identifying fake images on twitter during hurricane sandy,” in Proceedings of the 22nd international conference on World Wide Web, 2013, pp. 729–736.

\bibitem{castillo2011information} C. Castillo, M. Mendoza, and B. Poblete, “Information credibility on twitter,” in Proceedings of the 20th international conference on World wide web, 2011, pp. 675–684.

\bibitem{gupta2012twitter} A. Gupta and P. Kumaraguru, “Twitter explodes with activity in mumbai blasts! a lifeline or an unmonitored daemon in the lurking?,” 2012.

\bibitem{gupta2012credibility} A. Gupta and P. Kumaraguru, “Credibility ranking of tweets during high impact events,” in Proceedings of the 1st workshop on privacy and security in online social media, 2012, p. 2.
\
\bibitem{oh2011information} O. Oh, M. Agrawal, and H. R. Rao, “Information control and terrorism: Tracking the Mumbai terrorist attack through twitter,” Inf. Syst. Front., vol. 13, no. 1, pp. 33–43, 2011.

\bibitem{lee2014will} K. Lee, J. Mahmud, J. Chen, M. Zhou, and J. Nichols, “Who will retweet this?: Automatically identifying and engaging strangers on twitter to spread information,” in Proceedings of the 19th international conference on Intelligent User Interfaces, 2014, pp. 247–256.
\
\bibitem{lee2013warningbird} S. Lee and J. Kim, “Warningbird: A near real-time detection system for suspicious urls in twitter stream,” IEEE Trans. dependable Secur. Comput., vol. 10, no. 3, pp. 183–195, 2013.

\bibitem{gilani2017depth} Z. Gilani, R. Farahbakhsh, G. Tyson, L. Wang, and J. Crowcroft, “An in-depth characterisation of Bots and Humans on Twitter,” arXiv Prepr. arXiv1704.01508, 2017.
\
\bibitem{gilani2017classification} Z. Gilani, E. Kochmar, and J. Crowcroft, “Classification of twitter accounts into automated agents and human users,” in Proceedings of the 2017 IEEE/ACM International Conference on Advances in Social Networks Analysis and Mining 2017, 2017, pp. 489–496.

\bibitem{gilani2017bots} Z. Gilani, R. Farahbakhsh, G. Tyson, L. Wang, and J. Crowcroft, “Of bots and humans (on twitter),” in Proceedings of the 2017 IEEE/ACM International Conference on Advances in Social Networks Analysis and Mining 2017, 2017, pp. 349–354.

\bibitem{loria2014textblob} S. Loria et al., “Textblob: simplified text processing,” Second. TextBlob Simpl. Text Process., 2014.

\bibitem{wolfsfeld2013social} G. Wolfsfeld, E. Segev, and T. Sheafer, “Social media and the Arab Spring: Politics comes first,” Int. J. Press., vol. 18, no. 2, pp. 115–137, 2013.

\bibitem{morozov2014analysing} E. Morozov and M. Sen, “Analysing the Twitter social graph: Whom can we trust?,” MS thesis, Dept. Comput. Sci., Univ. Nice Sophia Antipolis, Nice, France, 2014.

\bibitem{alrubaian2017reputation} M. Alrubaian, M. Al-Qurishi, M. Al-Rakhami, M. M. Hassan, and A. Alamri, “Reputation-based credibility analysis of Twitter social network users,” Concurr. Comput. Pract. Exp., vol. 29, no. 7, p. e3873, 2017.

\bibitem{hughes2009twitter} A. L. Hughes and L. Palen, “Twitter adoption and use in mass convergence and emergency events,” Int. J. Emerg. Manag., vol. 6, no. 3–4, pp. 248–260, 2009.

\bibitem{Han} X. Han, X. Gu, and S. Peng, “Analysis of Tweet Form’s effect on users’ engagement on Twitter,” Cogent Bus. Manag., vol. 6, no. 1, pp. 1–15, Jan. 2019.

\bibitem{kang2012modeling} B. Kang, J. O’Donovan, and T. Höllerer, “Modeling topic specific credibility on twitter,” in Proceedings of the 2012 ACM international conference on Intelligent User Interfaces, 2012, pp. 179–188.

\bibitem{odonovan2012credibility} J. ODonovan, B. Kang, G. Meyer, T. Höllerer, and S. Adalii, “Credibility in context: An analysis of feature distributions in twitter,” in 2012 International Conference on Privacy, Security, Risk and Trust and 2012 International Confernece on Social Computing, 2012, pp. 293–301.

\bibitem{garcia2017understanding} D. Garcia, P. Mavrodiev, D. Casati, and F. Schweitzer, “Understanding popularity, reputation, and social influence in the twitter society,” Policy \& Internet, vol. 9, no. 3, pp. 343–364, 2017.

\bibitem{chen2011tweet} G. M. Chen, “Tweet this: A uses and gratifications perspective on how active Twitter use gratifies a need to connect with others,” Comput. Human Behav., vol. 27, no. 2, pp. 755–762, 2011.

\bibitem{amleshwaram2013cats} A. A. Amleshwaram, N. Reddy, S. Yadav, G. Gu, and C. Yang, “Cats: Characterizing automation of twitter spammers,” in 2013 Fifth International Conference on Communication Systems and Networks (COMSNETS), 2013, pp. 1–10.

\bibitem{yang2013empirical} C. Yang, R. Harkreader, and G. Gu, “Empirical evaluation and new design for fighting evolving twitter spammers,” IEEE Trans. Inf. Forensics Secur., vol. 8, no. 8, pp. 1280–1293, 2013.

\bibitem{fazil2018hybrid} M. Fazil and M. Abulaish, “A hybrid approach for detecting automated spammers in twitter,” IEEE Trans. Inf. Forensics Secur., vol. 13, no. 11, pp. 2707–2719, 2018.

\bibitem{mccoy2017university} C. G. McCoy, M. L. Nelson, and M. C. Weigle, “University Twitter engagement: using Twitter followers to rank universities,” arXiv Prepr. arXiv1708.05790, 2017.

\bibitem{leavitt2009influentials} A. Leavitt, E. Burchard, D. Fisher, and S. Gilbert, “The influentials: New approaches for analyzing influence on twitter,” Web Ecol. Proj., vol. 4, no. 2, pp. 1–18, 2009.

\bibitem{preussler2010managing} A. Preussler and M. Kerres, “Managing reputation by generating followers on Twitter,” Medien. Explor. Vis. und kollaborativer Wissensräume, pp. 129–143, 2010.

\bibitem{kerres2010managing} M. Kerres and A. Preussler, “Managing reputation by generating followers on Twitter.” Medien-Wissen-Bildung, 2010.

\bibitem{SocialMediaPopulation} J. Clement, Ed., “Number of social network users worldwide from 2010 to 2021 (in billions).” 2019.

\bibitem{Stefan} “A Beginner’s Guide to Active Learning - DataCamp.” [Online]. Available: https://www.datacamp.com/community/tutorials/active-learning. [Accessed: 28-Jul-2020].

\bibitem{Settles} B. Settles, “Computer Sciences Department Active Learning Literature Survey,” University of Wisconsin-Madison Department of Computer Sciences, 2009.

%

\bibitem{Noble} W. S. Noble, “What is a support vector machine?,” Nature Biotechnology, vol. 24, no. 12. Nature Publishing Group, pp. 1565–1567, Dec-2006.

\bibitem{Biau} G. Biau and G. B. Fr, “Analysis of a Random Forests Model,” 2012.

\bibitem{Danka} T. Danka and P. Horvath, “modAL: A modular active learning framework for Python,” May 2018.


\bibitem{Tong} D. K. Simon Tong, “Support vector machine active learning with applications to text classification,” J. Mach. Learn. Res., vol. 1, 2000.

\end{thebibliography}
\end{document}